\documentclass[reprint,superscriptaddress,amsmath, amssymb,aps,prd,notitlepage,longbibliography,floatfix,nofootinbib,onecolumn]{revtex4-1}

\usepackage{amsmath}
\usepackage{lipsum}
\allowdisplaybreaks[4]
\usepackage{cancel}
\usepackage{extarrows}
\usepackage{tensor}     
\usepackage{float}
\usepackage[caption = false]{subfig} 
\usepackage[final]{graphicx}   
\usepackage[
colorlinks=true,        
citecolor=blue,         
linkcolor=blue,         
urlcolor=blue           
]{hyperref}             
\usepackage{bm}         
\usepackage{xcolor}     
\usepackage{lipsum}
\usepackage{color}      
\usepackage[utf8]{inputenc} 
\usepackage[section]{placeins} 
\usepackage{appendix}
\usepackage{units}
\usepackage[capitalise]{cleveref}
\usepackage{units}

\usepackage{booktabs}
\usepackage{multirow}
\usepackage{import}
\usepackage{adjustbox}
\usepackage{url}
\newcommand{\nc}{\newcommand*}

\nc{\xbar}{\bar{x}}
\nc{\rhoeq}{\rho_{\mathrm{eq}}}
\nc{\zeq}{z_{\mathrm{eq}}}
\nc{\tla}{\tilde{\lambda}}
\nc{\bt}{\beta}
\nc{\dt}{\delta}
\nc{\Dt}{\Delta}
\nc{\vj}{\vec{j}}
\nc{\vl}{\vec{l}}
\nc{\hx}{\hat{x}}
\nc{\hy}{\hat{y}}
\nc{\bj}{\bm{j}}
\nc{\mJ}{\mathcal{J}}
\nc{\mP}{\mathcal{P}}
\nc{\Msun}{M_\odot}
\nc{\app}{\approx}
\nc{\av}[1]{\langle #1 \rangle}
\nc{\eq}[1]{Eq.~\eqref{#1}}
\nc{\al}{\alpha}
\nc{\Xstar}{X_{\ast}}
\nc{\fpbh}{f_{\mathrm{pbh}}}
\nc{\vth}{\vec{\theta}}
\nc{\vla}{\vec{\lambda}}
\nc{\vd}{\vec{d}}
\nc{\Mmin}{M_{\mathrm{min}}}
\nc{\rmd}{\mathrm{d}}
\nc{\mmin}{{m_{\mathrm{min}}}}
\nc{\mmax}{{m_{\mathrm{max}}}}
\nc{\mR}{\mathcal{R}}
\nc{\tmR}{\tilde{\mathcal{R}}}
\nc{\s}{\sigma}
\nc{\ogw}{\Omega_{\mathrm{GW}}}
\nc{\addref}{[\textcolor{red}{add ref}] }
\nc{\Om}{\Omega}
\nc{\gm}{\gamma}
\nc{\Gm}{\Gamma}
\nc{\gpcyr}{\mathrm{Gpc}^{-3}\,\mathrm{yr}^{-1}}
\nc{\Eq}[1]{Eq.~\eqref{#1}}
\nc{\Fig}[1]{Fig.~\ref{#1}}
\nc{\Table}[1]{Table~\ref{#1}}
\nc{\lvc}{LIGO/Virgo} 
\nc{\Sec}[1]{Sec.~\ref{#1}}
\nc{\eg}{\textit{e.g.~}}
\nc{\SNR}{\mathrm{SNR}}
\nc{\be}{\mathbf{\epsilon}}
\nc{\bn}{\mathbf{n}}
\nc{\bd}{\mathbf{d}}
\nc{\ba}{\mathbf{a}}
\nc{\eps}{\epsilon}
\nc{\bnu}{\mathbf{\nu}}
\nc{\mb}{\mathbf}
\nc{\bbt}{\mathbf{t}}
\nc{\bth}{\mathbf{\theta}}
\nc{\bep}{\mathbf{\epsilon}}
\nc{\uni}{\mathrm{U}}
\nc{\logu}{\operatorname{\mathrm{log-U}}}
\nc{\RN}{\mathrm{RN}}
\nc{\BN}{\mathrm{BN}}
\nc{\GN}{\mathrm{GN}}
\nc{\mcN}{\mathcal{N}}
\nc{\GWB}{\mathrm{GW}}
\nc{\yr}{\mathrm{yr}}
\nc{\Am}{\mathcal{A}}
\nc{\Dm}{\mathcal{D}}
\nc{\Hm}{\mathcal{H}}
\nc{\sovast}{Soviet Ast.}

\nc{\kmsmpc}{\mathrm{km\ s^{-1} Mpc^{-1}}}
\nc{\lcdm}{\Lambda\mathrm{CDM}}
\nc{\ev}{\mathrm{eV}}

\nc{\mrm}{\mathrm}
\nc{\BE}{B\scriptsize{AYES}\normalsize{E}\scriptsize{PHEM}\normalsize  }

\nc{\Ostgw}{\Omega_{\mathrm{GW}}^{\mathrm{ST}}}
\nc{\Ottgw}{\Omega_{\mathrm{GW}}^{\mathrm{TT}}}
\nc{\Ovlgw}{\Omega_{\mathrm{GW}}^{\mathrm{VL}}}
\nc{\Oslgw}{\Omega_{\mathrm{GW}}^{\mathrm{SL}}}
\nc{\cosxi}{\beta}

\nc{\gmPL}{\gamma_{\mathrm{PL}}}
\nc{\APL}{A_{\mathrm{PL}}}

\def\({\left(}
\def\){\right)}
\def\[{\left[}
\def\]{\right]}

\def\e{\begin{equation}}
\def\q{\end{equation}}
\def\m{\begin{eqnarray}}
\def\n{\end{eqnarray}}
\nc{\red}[1]{\textcolor{red}{#1}}

\begin{document}


\title{Constraints on Redshift-Binned Dark Energy using DESI BAO Data}

\author{Ye-Huang Pang}
\email{pangyehuang22@mails.ucas.ac.cn}
\affiliation{School of Fundamental Physics and Mathematical Sciences, Hangzhou Institute for Advanced Study, UCAS, Hangzhou 310024, China}
\affiliation{School of Physical Sciences,
    University of Chinese Academy of Sciences,
    No. 19A Yuquan Road, Beijing 100049, China}
\affiliation{CAS Key Laboratory of Theoretical Physics,
    Institute of Theoretical Physics, Chinese Academy of Sciences,Beijing 100190, China}
\author{Xue Zhang}
\email{corresponding author: zhangxue@yzu.edu.cn}
\affiliation{Center for Gravitation and Cosmology,
    College of Physical Science and Technology,
    Yangzhou University, Yangzhou 225009, China}
\author{Qing-Guo Huang}
\email{corresponding author: huangqg@itp.ac.cn}
\affiliation{School of Fundamental Physics and Mathematical Sciences, Hangzhou Institute for Advanced Study, UCAS, Hangzhou 310024, China}
\affiliation{School of Physical Sciences,
    University of Chinese Academy of Sciences,
    No. 19A Yuquan Road, Beijing 100049, China}
\affiliation{CAS Key Laboratory of Theoretical Physics,
    Institute of Theoretical Physics, Chinese Academy of Sciences,Beijing 100190, China}


\begin{abstract}

We parameterize the equation of state of late-time dark energy as $w_{\mathrm{bin}}(z)$, with three redshift bins, characterized by a constant equation of state in each bin. Then, we constrain the parameters of the $w_{\mathrm{bin}}$CDM model using datasets from DESI BAO data, \textit{Planck} CMB power spectrum, ACT DR6 lensing power spectrum, and type Ia supernova distance-redshift data of Pantheon Plus/DES Y5/Union3. The significances for $w_1>-1$ is $1.9\sigma$, $2.6\sigma$ and $3.3\sigma$, and $w_2$ is consistent with $-1$ within $1\sigma$ level, while $1.6\sigma$, $1.5\sigma$ and $1.5\sigma$ for $w_3<-1$ in these three data combinations with different choices of type Ia supernova datasets, respectively. Additionally, to alleviate $H_0$ tension, we incorporate the early dark energy (EDE) model in the early-time universe and add the SH0ES absolute magnitude $M_b$ prior (or $H_0$ prior) to further constrain the $w_{\mathrm{bin}}$EDE model. 
In the $w_{\mathrm{bin}}$EDE model, we find a $w_{\mathrm{bin}}(z)$ pattern similar to that in the $w_{\mathrm{bin}}$CDM model.
The results of the three data combinations exhibit $w_1>-1$ at $1.9\sigma$, $1.7\sigma$ and $2.9\sigma$ level, meanwhile $w_3<-1$ at  $1.3\sigma$, $1.3\sigma$ and $1.3\sigma$ level, respectively. In all, our results indicate that the transition of dark energy from phantom at high redshifts to quintessence at low redshifts is not conclusive. 

\end{abstract}
\maketitle

\section{Introduction}
The accelerated expansion of the universe at the low redshift remains a mystery, awaiting to be uncovered through more preciser observations. Various astrophysical observations have been dedicated to tracking clues that can shed light on the evolution of dark energy evolution, particularly its equation of state (EoS). Observations of Type Ia supernovae (SN Ia) \cite{Pan-STARRS1:2017jku, Brout:2022vxf} and large-scale structures \cite{Ross:2014qpa, BOSS:2016wmc} could potentially illuminate the characteristics of dark energy.

The first-year cosmological results from DESI (Dark Energy Spectroscopy Instrument) have been released \cite{DESI:2024uvr, DESI:2024lzq, DESI:2024mwx}. The results of the DESI 2024 BAO and their implications for dynamical dark energy continue to attract significant interest.
In Ref. \cite{DESI:2024mwx}, the constraints of the $w_0 w_a$CDM model exhibit a deviation from the $\Lambda$CDM model by $2.5\sigma$/$3.5\sigma$/$3.9\sigma$, when using the data combination of CMB+DESI BAO+Pantheon Plus, CMB+DESI BAO+Union3 and CMB+DESI BAO+DES Y5, respectively. Concerns regarding the anomalous data points and identifying those that may contribute to the constrained dynamical behavior of dark energy has been discussed in Ref. \cite{Colgain:2024xqj, Wang:2024pui}. Numerous constraints on dark energy model reveal implications for different dark energy dynamics using the DESI data \cite{Tada:2024znt, Berghaus:2024kra, Shlivko:2024llw, Bhattacharya:2024hep, Ramadan:2024kmn, Giare:2024smz, Escamilla-Rivera:2024sae, Gialamas:2024lyw}. Model-independent reconstructions of dark energy evolution also provide insights directly from the data points through Gaussian process \cite{Yang:2024kdo, Mukherjee:2024ryz} and crossing statistics \cite{DESI:2024aqx}. Refer to recent relevant works in Ref.  \cite{Du:2024pai,Ye:2024ywg,Li:2024qso,Wang:2024sgo,Liu:2024gfy,Liu:2024txl,Wang:2024hwd,Wang:2024rus,Huang:2024qno,Gu:2024jhl}.

Furthermore, there have been some discussions about the consistency with other cosmological constraints \cite{Bousis:2024rnb, Jia:2024wix, DES:2024ywx}. The DESI data provide a new independent determination of the value of $H_0$,  and Ref.  \cite{Pogosian:2024ykm} found that combining DESI BAO with priors on $\Omega_m h^2$ and angular scale of sound horizon at recombination $\theta_*$ yields a value of  $H_0 = 69.88\pm 0.93$ $\kmsmpc$, which is in tension with both \textit{Planck} ($H_0 = 67.27 \pm 0.60$ $\kmsmpc$) \cite{Planck:2018vyg} and SH0ES measurements ($H_0 = 73.04 \pm 1.04$ $\kmsmpc$) \cite{Riess:2021jrx}.
Various models aimed at alleviating the Hubble tension have also been constrained using the DESI data, including early dark energy (EDE) \cite{Wang:2024dka}, varying electron mass \cite{Seto:2024cgo}, etc.
It is argued that addressing both the $H_0$ tension and $S_8$ tension may involve late-time extensions, in addition to early-time new physics \cite{Reboucas:2023rjm, Vagnozzi:2023nrq}. Therefore, investigating constraints on late-time dark energy remains crucial for exploring cosmological concordance and assessing the feasibility of this hypothesis.

In this work, we employ a widely used parameterization for the dark energy EoS in the cosmological model \cite{SupernovaCosmologyProject:2008ojh, Huang:2009rf, Huang:2016fxc}, dividing redshifts into several bins with a constant EoS within each bin. Then, we utilize DESI BAO data and other observational data to constrain the cosmological parameters.
Additionally, we incorporate the EDE model and apply a binned parameterization for late-time dark energy to investigate the constraint results in this model, which is free from Hubble tension.
This paper is organized as follows. In \Sec{models}, we present the binned parameterization of dark energy equation of state and its corresponding models. The data combinations and analysis methods are detailed in \Sec{data}. The results and related discussions are presented in \Sec{results}. Finally, we conclude in \Sec{conclusions}.

\section{Models}\label{models}
In the late-time of the universe, in order to comprehend the evolution of EoS, we consider the $w_{\mathrm{bin}}$ dark energy. The redshift binned dark energy EoS parameterization in our study is artificially selected based on
\begin{equation}
    w_{\mathrm{bin}}(z) =
    \begin{cases}w_1, & 0\leqslant z< 0.4 \\
    w_2, & 0.4\leqslant z < 0.8 \\
    w_3, & 0.8 \leqslant z<2.1 \\
    -1, & z\geqslant 2.1
    \end{cases}.
\end{equation}
The parameterization of dark energy EoS is non-continuous, with $w_1, w_2$ and $w_3$ being constant. In the redshift range $z>2.1$, in light of insufficient data, we assume the $\Lambda$CDM model. The EoS constraint is primarily influenced by the data in the respective redshift bin, providing additional complementary results for probing the dark energy.
We employ the $w_{\mathrm{bin}}$ dark energy in place of the cosmological constant in $\Lambda$CDM model, resulting a new model is referred to as $w_{\mathrm{bin}}$CDM model.

To alleviate the $H_0$ tension, in addition to the standard early-time cosmological evolution, we also consider the early dark energy model \cite{Poulin:2018cxd} combinned with late-time dark energy described by $w_{\mathrm{bin}}(z)$, which is labeled as the $w_{\mathrm{bin}}$EDE model.
The early dark energy model introduces a scalar field $\phi$ with an axion-like potential,
\begin{equation} \label{V_phi}
    V(\phi) = m^2 f^2 [1-\cos(\phi/f)]^n,
\end{equation}
which becomes activate around recombination. This reduces the sound horizon of recombination epoch and thus increases the inferred value of $H_0$ from the model fitting. Similar to the axion potential when $n=1$, $m$ represents the mass of the scalar field, and $f$ is the axion decay constant. Generally, $n$ is chosen to be 3. This scenario has been extensively discussed in previous research \cite{Hill:2020osr,Murgia:2020ryi,Ivanov:2020ril,Poulin:2023lkg}. The parameters related to the EDE scalar field are $\{\log_{10}a_c, f_{\mathrm{EDE}}, \phi_i\}$, where $a_c$ corresponds to the scale factor when EDE reaches its maximum fractional energy density, $f_{\mathrm{EDE}}$ represents the fractional energy density of EDE at $a_c$, and $\phi_i$ denotes the initial value of the scalar field.

\section{Datasets and Methodology}\label{data}
To constrain the $w_{\mathrm{bin}}$CDM model mentioned above, we employ a combination of CMB, DESI BAO, and SN datasets as in \cite{DESI:2024mwx}.
\begin{itemize}
    \item The CMB data includes low-$\ell$ TT and EE power spectrum from \textit{Planck}, the high-$\ell$ TTTEEE power spectrum from PR3 \texttt{plik} likelihood \cite{Planck:2018lbu, Planck:2019nip}, and the CMB lensing power spectrum from ACT DR4 \cite{ACT:2023dou}.
    \item The DESI BAO data consists of 12 BAO samples obtained from observations on various tracers, including BGS ($z_{\mathrm{eff}} = 0.30$), LRG ($z_{\mathrm{eff}} = 0.51, 0.71$), ELG ($z_{\mathrm{eff}} = 1.32$), LRG + ELG ($z_{\mathrm{eff}} = 0.93$), quasar ($z_{\mathrm{eff}} = 1.49$), and Ly$\alpha$ forest ($z_{\mathrm{eff}} = 2.33$) \cite{DESI:2024mwx}.
    \item The SN data comprises non-calibrated supernova luminosity-distance measurements from Pantheon Plus ($0.001<z<2.26$) \cite{Brout:2022vxf}, Union3 \cite{Rubin:2023ovl}, or DES Y5 ($0.025<z<1.3$) \cite{DES:2024tys}.
\end{itemize}

Due to the sampling prior volume effect \cite{Herold:2021ksg, Poulin:2023lkg, Efstathiou:2023fbn}, it is important to include the SH0ES $H_0$ prior when sampling the $w_{\mathrm{bin}}$EDE model. Regardless of the discussion on the biased $H_0$ posterior \cite{Hill:2020osr, DiValentino:2021izs}, this $H_0$ prior is indeed necessary for new physics to alleviate the Hubble tension. Therefore, the data combination used to constrain $w_{\mathrm{bin}}$EDE model is CMB+DESI BAO+SN+SH0ES. When SN data is Pantheon Plus, we use a prior on supernova absolute magnitude $M_b$ \cite{Riess:2021jrx} instead of $H_0$ prior, and in this case,
the results with these two priors are almost identical \cite{Camarena:2021jlr, Efstathiou:2021ocp}.
We adopt flat prior on $w_1, w_2$ and $w_3$ over range $[-3, 1]$.

The cosmological parameter constraints is inferred through the chains of MCMC (Markov-chain Monte Carlo) sampling.
We use the MCMC sampler of \texttt{cobaya} \cite{Torrado:2020dgo} and a modified version of \texttt{class} \cite{Blas:2011rf, Murgia:2020ryi} \footnote{The modified version of \texttt{class} for EDE can be assessed at \url{https://github.com/PoulinV/AxiCLASS}.} to calculate observable quantities  .
The MCMC chains are analysed using \texttt{getdist} \cite{Lewis:2019xzd}.
To ensure convergence in our analysis, we require the Gelman-Rubin criterion \cite{Gelman:1992zz} $R-1<0.05$ for sampling process.

\section{Results and Discussion} \label{results}
\begin{figure}[!b]
    \centering
    \includegraphics[width=\linewidth]{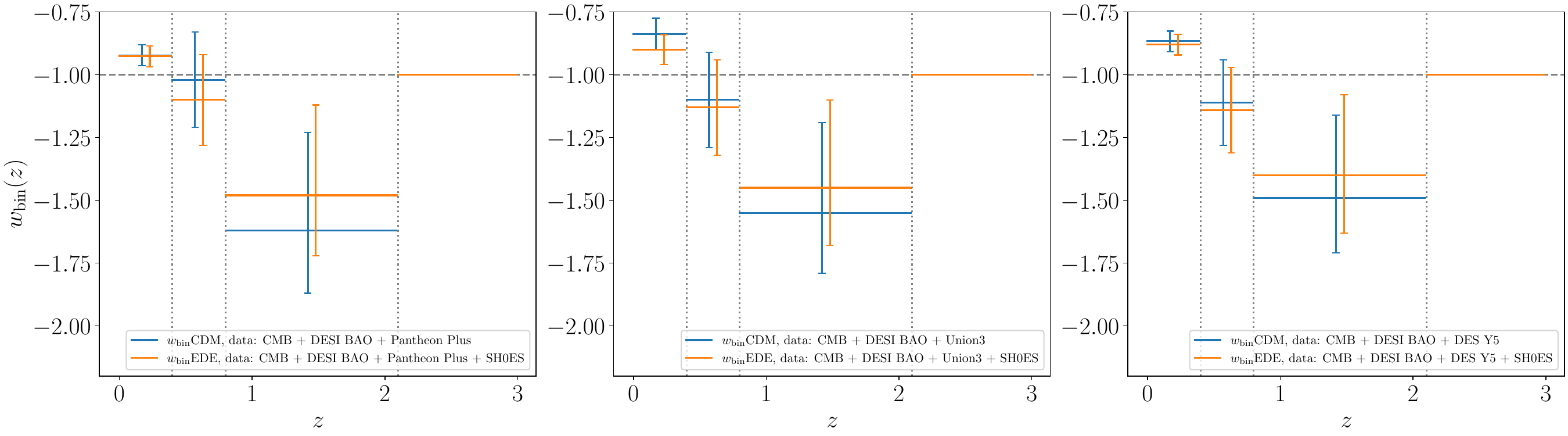}
    \caption{The mean and $1\sigma$ confidence level (C.L.) of late-time dark energy EoS constraints with a combination of CMB + DESI BAO + SN data for the $w_{\mathrm{bin}}$CDM model (blue) and the $w_{\mathrm{bin}}$EDE model (orange). The SN data in the combined dataset correspond to Pantheon Plus, DES Y5, and Union3 from left to right. For clarity, we have offset the error bar plots of $w_i$, where $i = 1,2,3$.}
    \label{fig:wbin}
\end{figure}

\begin{figure}[b]
    \centering
        \includegraphics[width=0.45\textwidth]{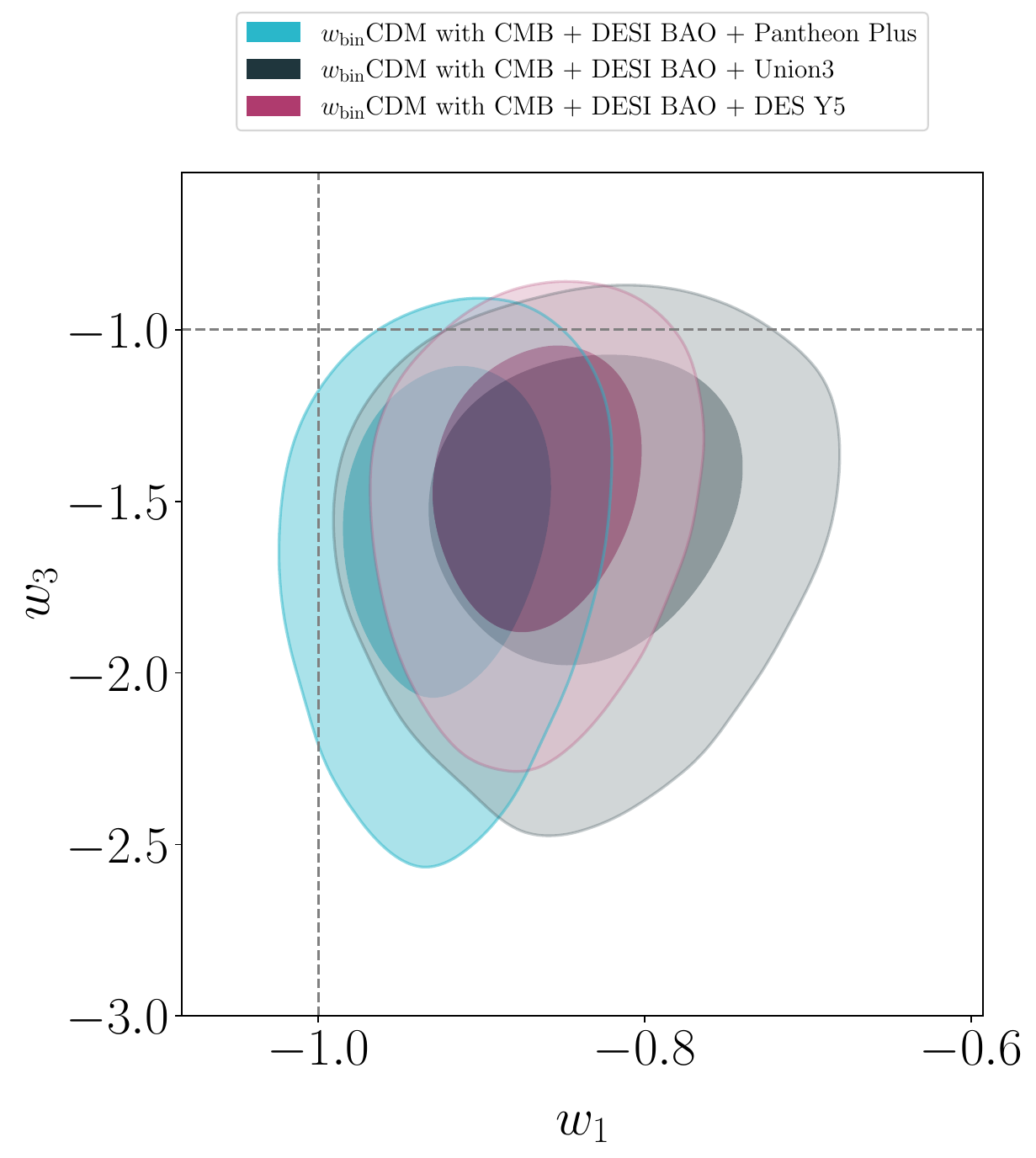}
        \includegraphics[width=0.43\textwidth]{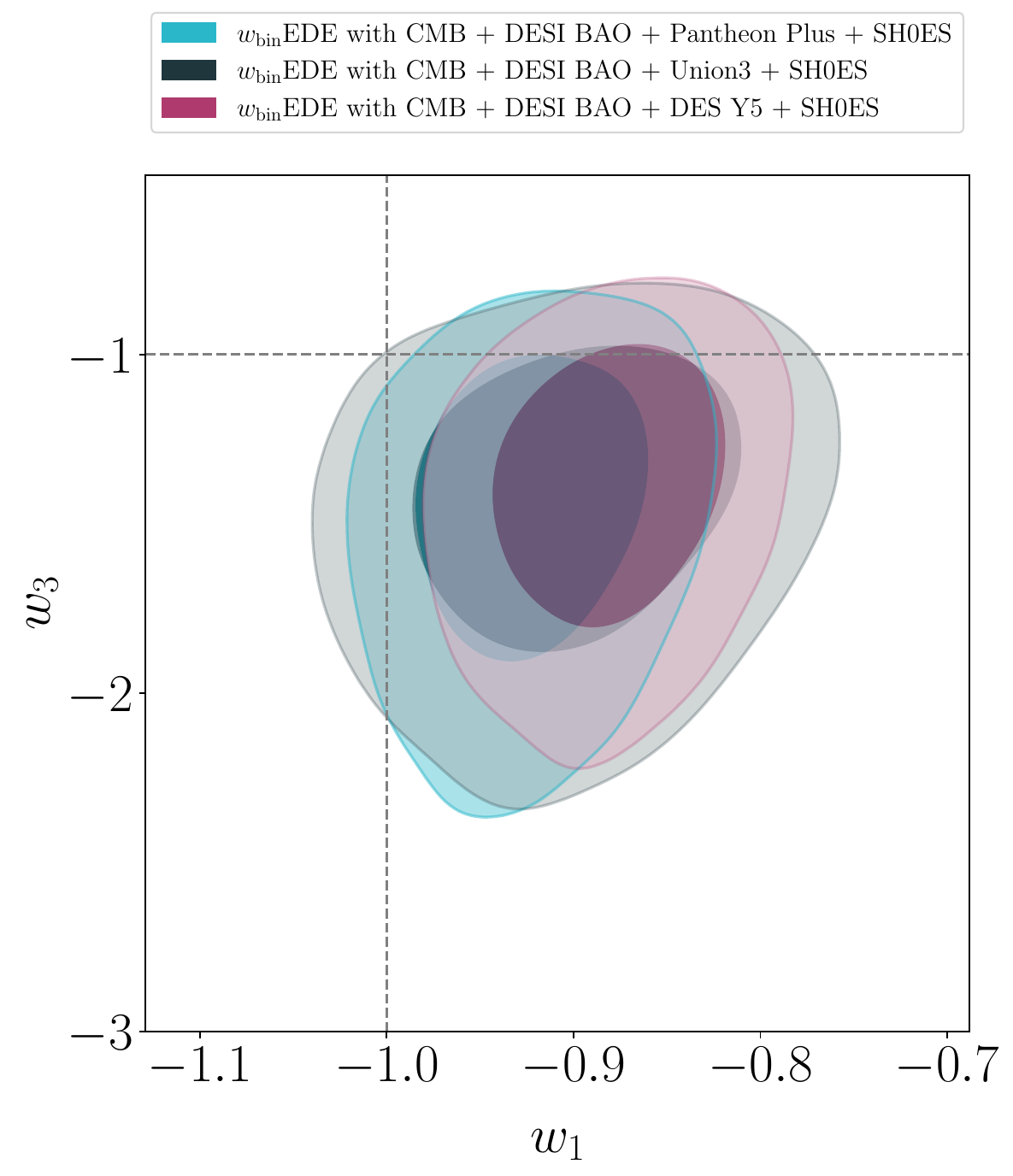}
    \caption{Marginalized posterior distributions for $w_1$ and $w_3$ in the $w_{\mathrm{bin}}$CDM model (left panel) and $w_{\mathrm{bin}}$EDE model (right panel).}
    \label{fig:w1w3}
\end{figure}

\begin{figure}[hp]
    \centering
    \includegraphics[width=\linewidth]{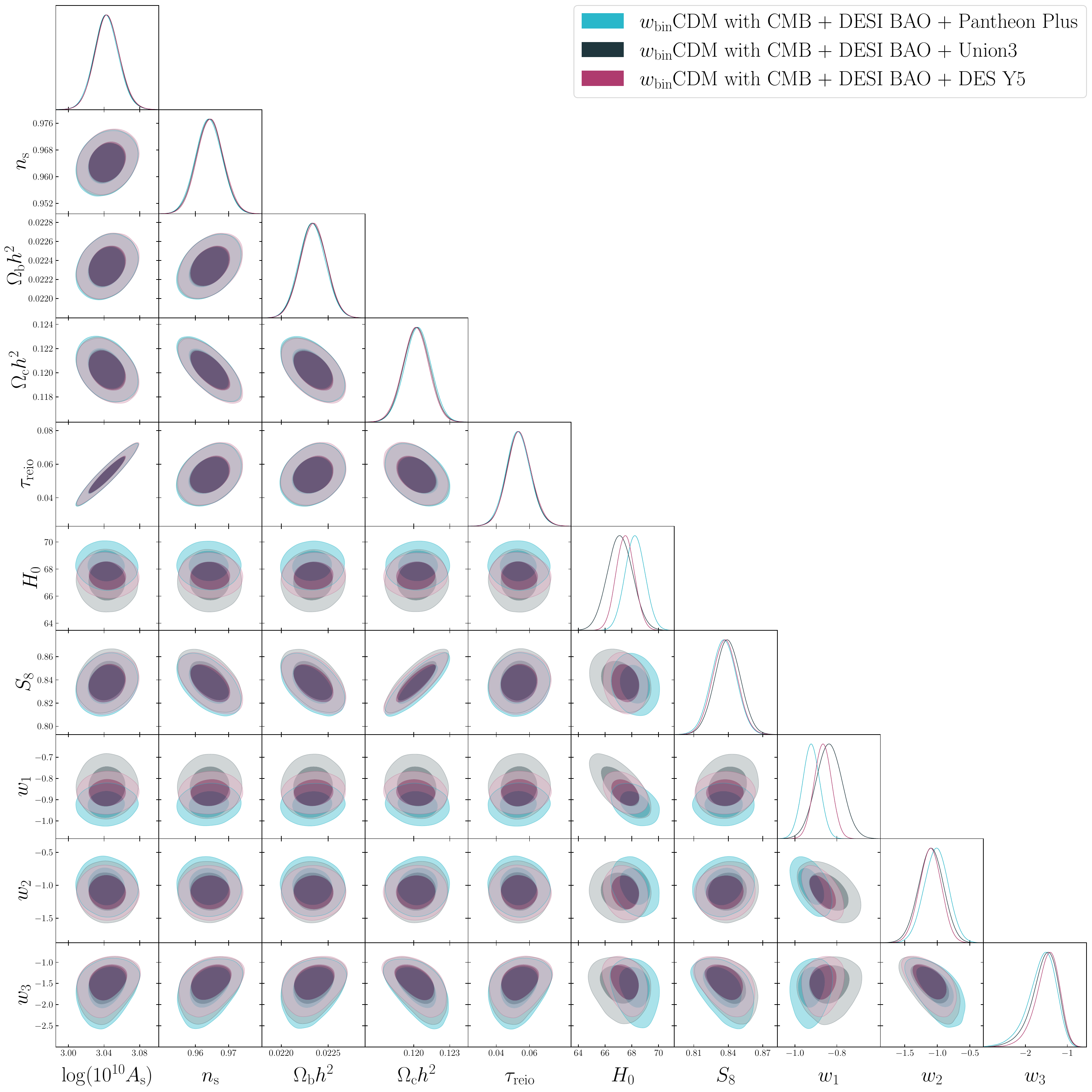}
    \caption{The 2D posterior distribution of the $w_{\mathrm{bin}}$CDM model is presented at $1\sigma$ and $2\sigma$ C.L. The parameters are constrained using a combination of CMB + DESI BAO + Pantheon Plus/Union3/DES Y5 data.}
    \label{fig:wbincdm}
\end{figure}

\begin{figure}[hp]
    \centering
    \includegraphics[width=\linewidth]{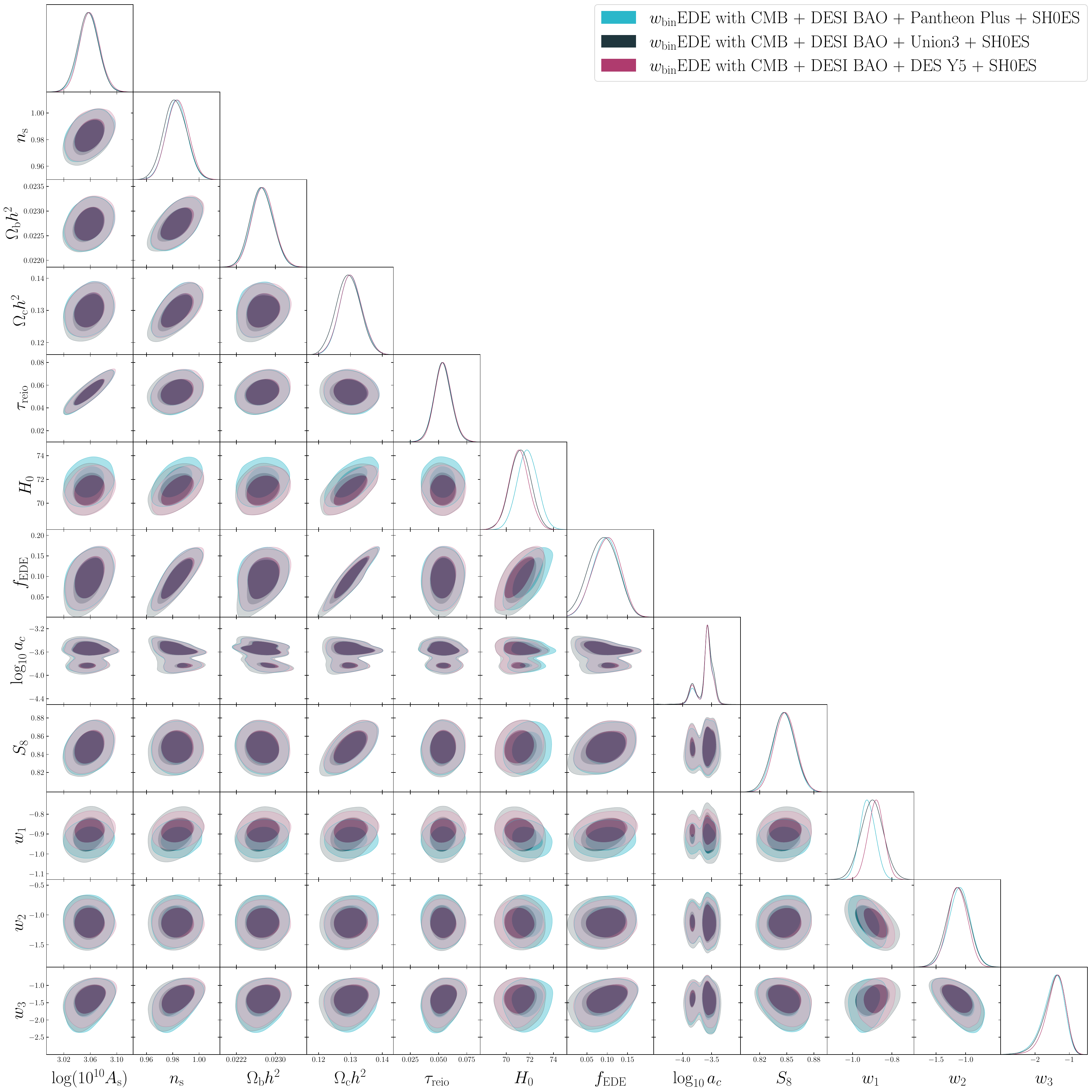}
    \caption{Th 2D posterior distribution of the $w_{\mathrm{bin}}$EDE model is shown at $1\sigma$ and $2\sigma$ C.L. The parameter constraints are derived from a combination of CMB + DESI BAO + Pantheon Plus/Union3/DES Y5 + SH0ES data.}
    \label{fig:wbinede}
\end{figure}

In \Fig{fig:wbin}, we present the mean value and the corresponding $1\sigma$ C.L. of $w_i~(i=1, 2, 3)$ in each redshift bin inferred from MCMC chains. 
Fig. \ref{fig:w1w3} shows the $w_1-w_3$ plane in the $w_{\mathrm{bin}}$CDM and $w_{\mathrm{bin}}$EDE using different data combinations.
The 2-dimensional posterior distribution of the $w_{\mathrm{bin}}$CDM and $w_{\mathrm{bin}}$EDE parameters is displayed in \Fig{fig:wbincdm} and \Fig{fig:wbinede} respectively. Additionally, we list the mean values and $1\sigma$ C.L. of the $w_{\mathrm{bin}}$CDM and $w_{\mathrm{bin}}$EDE parameters, along with the $\chi^2$ values for the best-fit parameters in Table \ref{tab:table1} and Table \ref{tab:table2}.

For the $w_{\mathrm{bin}}$CDM model, the EoS constraints in each redshift bin are $w_1 = -0.923 \pm 0.041$, $w_2 = -1.02 \pm 0.19 $ and $w_3 = -1.62^{+0.39}_{-0.25}$ at $1\sigma$ C.L. when using data combination of CMB+BAO+Pantheon Plus. We find that, in this case, $w_1>-1$ at $\sim 1.9 \sigma$ level, $w_2$ is consistent with $-1$ within the $1\sigma$ level and $w_3 <-1$ at $\sim 1.6 \sigma$ level. Although the mean $w_3$ value seems to deviate significantly from $-1$, the large error bar diminishes the statistical significance of this deviation. 
As shown in \Fig{fig:wbin}, CMB+BAO+Union3 and CMB+BAO+DES Y5 yield similar $w_{\mathrm{bin}}$ constraints to the CMB+BAO+Pantheon Plus case, with slightly different mean values of $w_i$ for $i = 1,2,3$. 
The significance of the tension with $\Lambda$CDM ($w=-1$) is $2.6\sigma$ and $3.3\sigma$ for $w_1$, while $1.5\sigma$ and $1.5\sigma$ for $w_3$ in these two data combinations respectively.
The detailed $w_i$ mean values and corresponding $1\sigma$ confidence levels are listed in Table \ref{tab:table1}. A comparison of the marginalized posterior distributions in $w_{\mathrm{bin}}$CDM model constrained by CMB+BAO+Pantheon Plus/Union3/DES Y5 can be seen in \Fig{fig:wbincdm}.
We find that CMB+BAO+Pantheon Plus/Union3/DES Y5 constraints are compatible with each other.

Compared to the model-independent reconstruction of $w(z)$, such as the crossing statistics reconstruction \cite{DESI:2024aqx}, our results for $w_{\mathrm{bin}}$ can be interpreted as flattening out the evolution of $w(z)$ with respect to redshift, making it a constant value within each redshift bin.
In Ref. \cite{DESI:2024aqx}, when reconstructed with CMB+BAO+Pantheon Plus/Union3/DES Y5 dataset, the redshift range $0<z\lesssim 0.5$ exhibits $w>-1$, while the range $1.0\lesssim z\lesssim 2.5$ shows $w<-1$ at $\gtrsim 1\sigma$ level. 
Therefore, it is not surprising that we observe our constrained results for $w_{\mathrm{bin}}$, indicating that 
mean values of $w_1$ lie in quintessence regime and mean values of $w_3$ lie in phantom regime at $> 1\sigma$ level.
However, due to the poor statistical significance, we cannot draw a simple conclusion about whether the data prefer the transition from phantom dark energy to quintessence. In the left panel of \Fig{fig:w1w3}, we show the 2-dimensional (2D) marginalized posterior distributions in $(w_1,w_3)$ plane. Since $w_2$ is consistent with $-1$ within the $1\sigma$ level, the $(w_1, w_3)$ contours provide a visual method to determine whether dark energy evolves from the phantom regime in the third redshift bin to the quintessence regime in the first redshift bin. Specifically, if the $(w_1, w_3)$ contour falls within the lower-right quadrant formed by the intersection of lines $w_1 = -1$ and $w_3 = -1$, it suggests that the dark energy EoS may exhibit phantom-like behavior at early times and be quintessence-like behavior at lower redshift. However, all of the $(w_1, w_3)$ contours constrained by CMB+DESI BAO+Pantheon Plus/Union3/DES Y5 intersect with line $w_3 = -1$, indicating that there is no significant preference for phantom dark energy at the third redshift bin.


For the $w_{\mathrm{bin}}$EDE model, the mean values and $1\sigma$ C.L. of EoS in each reshift bin are $w_1 = -0.925\pm 0.040$, $w_2 = -1.10\pm 0.18 $ and $w_3 = -1.48^{+0.37}_{-0.24}$ when using data combination of CMB+BAO+Pantheon Plus+SH0ES.
We find a similar trend, $w_1>-1$ at $\sim 1.9 \sigma$ level, $w_2$ is consistent with $-1$ within $1\sigma$ level, and $w_3<-1$ at $\sim 1.3 \sigma$ level. The corresponding constraints on $w_{\mathrm{bin}}$ from CMB+BAO+Union3+SH0ES and CMB+BAO+DES Y5+SH0ES are depicted in \Fig{fig:wbin}.
The results of these data combinations exhibit slightly discrepancies with $\Lambda$CDM at $1.7\sigma$ and $2.9\sigma$ level for $w_1$, meanwhile $1.3\sigma$ and $1.3\sigma$ level for $w_3$ respectively. 
The 2D marginalized posterior distributions of the $w_{\mathrm{bin}}$EDE model, constrained by CMB+BAO+Pantheon Plus/Union3/DES Y5+SH0ES, are illustrated in \Fig{fig:wbinede}.
These constraints demonstrate mutual compatibility. Additionally, the $(w_1, w_3)$ contours are plotted in the right panel of \Fig{fig:w1w3}, and the corresponding results do not prominently exhibit phantom to quintessence transition.

The fit to CMB+BAO+Pantheon Plus+SH0ES data yields $H_0 = 71.79\pm 0.86 \kmsmpc$ within the framework of the $w_{\mathrm{bin}}$EDE model, which is consistent with the SH0ES result of $H_0 = 73.04\pm 1.04 \kmsmpc$ at $1\sigma$ level. For the CMB+BAO+Union3+SH0ES data, the resulting value of $H_0 = 71.20\pm 0.88 \kmsmpc$ deviates by approximately $\sim 1.4\sigma$ from the SH0ES result. The combination of CMB+BAO+DES Y5+SH0ES data yields $H_0 = 71.08\pm 0.85 \kmsmpc$, which deviates from SH0ES by $\sim 1.5\sigma$.
In the EDE model, an increase of $H_0$ is accompanied by higher values of $n_s, \omega_c$, and an exacerbated $S_8$ tension compared to the $\Lambda$CDM model \cite{Ye:2021nej, Vagnozzi:2021gjh, Poulin:2023lkg}. Similar relative changes in these parameters are found when comparing constraints on $w_{\mathrm{bin}}$CDM parameters in \Fig{fig:wbincdm} with those on $w_{\mathrm{bin}}$EDE in \Fig{fig:wbinede}.
We find that early-time physics, tightly constrained by CMB data, subsequently impact constraints on $w_{\mathrm{bin}}$ within the $w_{\mathrm{bin}}$EDE model through shifts in parameters such as $\{ \log(10^{10} A_\mathrm{s}), n_\mathrm{s}, \Omega_\mathrm{b} h^2, \Omega_\mathrm{c} h^2, \tau_\mathrm{reio} \}$.




\begin{table}[!htbp]
    \centering
    \begin{tabular} { l  c  c   c}
        \hline
          & \ CMB+BAO+Pantheon Plus\  &\ CMB+BAO+Union3\  &\ CMB+BAO+DES Y5\ \\
        \hline
        {$\log(10^{10} A_\mathrm{s})$}
        & $3.042(3.041)\pm 0.014            $
        & $3.043(3.041)\pm 0.014            $
        & $3.044(3.043)\pm 0.014            $\\

        {$n_\mathrm{s}   $}
        & $0.9640(0.9634)\pm 0.0040          $
        & $0.9643(0.9634)\pm 0.0040          $
        & $0.9645(0.9654)\pm 0.0040          $\\

        {$H_0 [\kmsmpc]             $}
        & $68.24(67.995)\pm 0.74             $
        & $67.12(67.06)\pm 0.95             $
        & $67.51(67.36)\pm 0.69             $\\

        {$\Omega_\mathrm{b} h^2$}
        & $0.02233(0.02228)\pm 0.00014        $
        & $0.02234(0.02240)\pm 0.00014        $
        & $0.02234(0.02236)\pm 0.00014        $\\

        {$\Omega_\mathrm{c} h^2$}
        & $0.1203(0.1206)\pm 0.0011          $
        & $0.1202(0.1203)\pm 0.0011          $
        & $0.1202(0.1200)\pm 0.0011          $\\

        {$\tau_\mathrm{reio}$}
        & $0.0532(0.0525)\pm 0.0075          $
        & $0.0537(0.0524)\pm 0.0074          $
        & $0.0538(0.0527)\pm 0.0075          $\\

        {$w_{1}$}
        & $-0.923(-0.924)\pm 0.041           $
        & $-0.837(-0.827)\pm 0.062           $
        & $-0.866(-0.864)\pm 0.041           $\\

        {$w_{2}$}
        & $-1.02(-1.04)\pm 0.19             $
        & $-1.10(-1.14)\pm 0.19             $
        & $-1.11(-1.11)\pm 0.17             $\\

        {$w_{3}$}
        & $-1.62(-1.49)^{+0.39}_{-0.25}     $
        & $-1.55(-1.46)^{+0.36}_{-0.24}     $
        & $-1.49(-1.40)^{+0.33}_{-0.22}     $\\

        $S_8                       $
        & $0.836(0.839)\pm 0.011            $
        & $0.839(0.838)\pm 0.011            $
        & $0.837(0.836)\pm 0.011            $\\
        \hline
        {$\chi^2_{\mathrm{CMB}}$}
        & $1015.15$
        & $1015.22$
        & $1015.01$\\


        {$\chi^2_{\mathrm{SN}}$}
        & $1402.26$
        & $22.87$
        & $1639.31$\\

        {$\chi^2_{\mathrm{DESI\ BAO}}$}
        & $12.57$
        & $11.52$
        & $11.64$\\

        {$\chi^2_{\mathrm{bestfit}}$}
        & $2429.98$
        & $1049.61$
        & $2665.96$\\
        \hline
        \end{tabular}
    \caption{The mean (best-fit) and $\pm 1\sigma$ C.L. of the $w_{\mathrm{bin}}$CDM model parameters are shown in the table, along with the corresponding $\chi^2$ values  for the best-fit parameters. }
    \label{tab:table1}
\end{table}

\begin{table}[!htbp]
    \centering
    \begin{tabular} { l  c  c   c}
        \hline
          &  \ CMB+BAO+Pantheon Plus+SH0ES\  &\ CMB+BAO+Union3+SH0ES\  &\ CMB+BAO+DES Y5+SH0ES\ \\
        \hline
        {$\log(10^{10} A_\mathrm{s})$}
        & $3.058(3.066)\pm 0.016            $
        & $3.057(3.064)\pm 0.016            $
        & $3.059(3.074)\pm 0.015            $\\

        {$n_\mathrm{s}   $}
        & $0.9827(0.9860)\pm 0.0080          $
        & $0.9818(0.9872)\pm 0.0085          $
        & $0.9836(0.9895)\pm 0.0080          $\\

        {$H_0 [\kmsmpc]           $}
        & $71.79(72.21)\pm 0.86             $
        & $71.20(71.60)\pm 0.88             $
        & $71.08(71.70)\pm 0.85             $\\

        {$\Omega_\mathrm{b} h^2$}
        & $0.02272(0.02264)\pm 0.00022        $
        & $0.02272(0.02269)\pm 0.00023        $
        & $0.02274(0.02281)\pm 0.00023        $\\

        {$\Omega_\mathrm{c} h^2$}
        & $0.1300(0.1335)\pm 0.0035          $
        & $0.1294(0.1329)\pm 0.0037          $
        & $0.1302(0.1333)\pm 0.0036          $\\

        {$\tau_\mathrm{reio}$}
        & $0.0536(0.0549)\pm 0.0078          $
        & $0.0537(0.0558)\pm 0.0078          $
        & $0.0542(0.0595)\pm 0.0077          $
        \\

        {$f_{\mathrm{EDE}}$}
        & $0.096(0.127)\pm 0.032            $
        & $0.090(0.125)\pm 0.034            $
        & $0.099(1.312)\pm 0.033            $
        \\

        {$\log_{10}a_c   $}
        & $-3.61(-3.56)^{+0.15}_{-0.033}    $
        & $-3.61(-3.57)^{+0.18}_{-0.035}    $
        & $-3.60(-3.58)^{+0.14}_{-0.037}    $
        \\

        {$w_{1}$}
        & $-0.925(-0.914)\pm 0.040           $
        & $-0.901(-0.857)\pm 0.058           $
        & $-0.880(-0.878)\pm 0.041           $
        \\

        {$w_{2}$}
        & $-1.10(-1.18)\pm 0.18             $
        & $-1.13(-1.20)\pm 0.19             $
        & $-1.14(-1.14)\pm 0.17             $
        \\

        {$w_{3}$}
        & $-1.48(-1.20)^{+0.37}_{-0.24}     $
        & $-1.45(-1.26)^{+0.35}_{-0.23}     $
        & $-1.40(-1.20)^{+0.32}_{-0.22}     $
        \\

        $S_8                       $
        & $0.847(0.853)\pm 0.012            $
        & $0.847(0.851)\pm 0.013            $
        & $0.849(0.853)\pm 0.012            $
        \\

        \hline
        {$\chi^2_{\mathrm{CMB}}$}
        & $1014.45$
        & $1013.83$
        & $1015.42$\\

        {$\chi^2_{\mathrm{SN} + M_b}$}
        & $1454.12$
        & -
        & -\\

        {$\chi^2_{\mathrm{SN}}$}
        & -
        & $23.50$
        & $1639.77$\\

        {$\chi^2_{\mathrm{DESI\ BAO}}$}
        & $12.31$
        & $12.37$
        & $11.79$\\

        {$\chi^2_{\mathrm{bestfit}}$}
        & $2480.88$
        & $1053.53$
        & $2670.55$\\
        \hline
    \end{tabular}
    \caption{The mean (best-fit) and $\pm 1\sigma$ C.L. of $w_{\mathrm{bin}}$EDE model parameters are presented. The table also includes the $\chi^2$ values corresponding to the best-fit parameters.}
    \label{tab:table2}
\end{table}

\section{Summary and Conclusions}\label{conclusions}
In this paper, we examine the constraints on dark energy models using DESI 2024 BAO data and other cosmological dataset. We assume that the dark energy equation of state is constant in three redshift bins, with the equation of state being varied separately in each bin. We constrain the $w_{\mathrm{bin}}$CDM model and $w_{\mathrm{bin}}$EDE model using the \textit{Planck} CMB power spectrum, ACT lensing power spectrum, DESI BAO data, and Pantheon Plus / DES Y5 / Union3 data. Furthermore, we incorporate SH0ES prior into the combined datasets when sampling the $w_{\mathrm{bin}}$EDE model.

For the $w_{\mathrm{bin}}$CDM model, in the first redshift bin ($0\leqslant z<0.4$), the mean value of $w_{\mathrm{bin}}$, denoted as $w_1$, is greater than $-1$ and deviates from $\Lambda$CDM at $1.9\sigma$, $2.6\sigma$ and $3.3\sigma$ level. In the second redshift bin ($0.4\leqslant z<0.8$), $w_2$ is consistent with $-1$ within $1\sigma$ level. In the third redshift bin  ($0.8\leqslant z<2.1$), the mean value of $w_3$ less than $-1$ and deviates from $-1$ at $1.6\sigma$, $1.5\sigma$ and $1.5\sigma$ level. 
In addition to the $< 2\sigma$ deviation from cosmological constant-like EoS in the third redshift bin, our contour plots for $(w_1,w_3)$ show limited significance for early phantom-like dark energy behaviour.
Both the $w_{\mathrm{bin}}$CDM model and the $w_{\mathrm{bin}}$EDE model exhibit this characteristic when analyzed with their respective data combinations.
Although the mean values of $w_i$ between the two models are dissimilar to some extent, they remain consistent within $1\sigma$.
The observed variation in $w_i$ may be attributed to shifted model parameter values in the $w_{\mathrm{bin}}$EDE model compared to those in the $w_{\mathrm{bin}}$CDM model, as well as to distinct SN datasets employed in our analysis.
The incorporation of the $w_{\mathrm{bin}}$ parameterization into the EDE model does not compromise its ability to alleviate the $H_0$ tension. The resulting constraints on $H_0$, using CMB + DESI BAO + Pantheon Plus + SH0ES, are consistent with SH0ES measurements within $1\sigma$ level. For both the combinations of CMB + DESI BAO + Union3 + SH0ES and CMB + DESI BAO + DES Y5 + SH0ES, these tensions are at $<2\sigma$ confidence level.

\textit{Acknowledgements.}
We acknowledge the use of HPC Cluster of ITP-CAS. XZ is supported by grant from NSFC (Grant No. 12005183). QGH is supported by the grants from NSFC (Grant No.~12475065, 11991052), Key Research Program of Frontier Sciences, CAS, Grant No.~ZDBS-LY-7009, and China Manned Space Program through its Space Application System.

\bibliography{refs}
\end{document}